\documentclass[aps,prd,onecolumn,eqsecnum,amsmath,nofootinbib,preprintnumbers]{revtex4}%

\usepackage{color,graphicx,float,subfigure}
\usepackage{amsfonts,amssymb,theorem,mathrsfs,times}
\usepackage{bm}
\usepackage{mathtools}
\usepackage{amsfonts,amssymb,theorem,mathrsfs}
\usepackage{dsfont}
\usepackage{setspace}
\usepackage{graphicx}
\usepackage{ulem}
\usepackage[colorlinks,linkcolor=blue,anchorcolor=blue,citecolor=blue]{hyperref}
{\theorembodyfont{\upshape}
}
{\theorembodyfont{\upshape}
}
{\theorembodyfont{\upshape}
}
{\theorembodyfont{\upshape}
}
{\theorembodyfont{\upshape}
}
{\theorembodyfont{\upshape}
}

\newcommand{\dalm}{\kern1pt\vbox{\hrule height 0.9pt\hbox{\vrule width
0.9pt\hskip 2.5pt\vbox{\vskip 5.5pt}\hskip 3pt\vrule width
0.3pt}\hrule height 0.3pt}\kern1pt}

\begin{document}
\preprint{\hfill {\small{ICTS-USTC/PCFT-23-23}}}
\title{The instability of the inner horizon of the quantum-corrected black hole}

%

\author{ Li-Ming Cao$^{a\, ,b}$\footnote{e-mail
address: caolm@ustc.edu.cn}}

\author{ Long-Yue Li$^b$\footnote{e-mail
address: lily26@mail.ustc.edu.cn}}

\author{ Liang-Bi Wu$^b$\footnote{e-mail
address: liangbi@mail.ustc.edu.cn}}

\author{ Yu-Sen Zhou$^b$\footnote{e-mail
address: zhou\_ys@mail.ustc.edu.cn}}

\affiliation{$^a$Peng Huanwu Center for Fundamental Theory, Hefei, Anhui 230026, China}

\affiliation{${}^b$
Interdisciplinary Center for Theoretical Study and Department of Modern Physics,\\
University of Science and Technology of China, Hefei, Anhui 230026,
China}


\date{\today}


\begin{abstract}

We analyse the stability of the inner horizon of the  quantum-corrected black hole which is proposed in loop quantum gravity as the exterior of the quantum Oppenheimer-Snyder and Swiss Cheese models.
It is shown that the flux and energy density of a test scalar field measured by free-falling observers are both divergent near the Cauchy horizon.
By considering the generalized Dray-'t Hooft-Redmond  relation which is independent of the field equation, we find that the mass inflation always happens and  the scalar curvature and Kretschmann scalar are also divergent on the inner horizon.
These suggest that the inner horizon is unstable and will probably turn into a null singularity.
The results support the strong cosmic censorship hypothesis.
However, this also implies that the quantum corrected  model may not be the definitive endpoint as a regular black hole.
Besides, it further proposes that it may be challenging to observe the astronomical phenomenon which depends on the existence of the inner horizon of the black hole.

\end{abstract}


\maketitle


\section{Introduction}
Up to date, people believe that General relativity (GR) is the most successful theory for describing gravity.
One of the most significant predictions of GR is the existence of black holes  which contain singularities.
Physicists attempt to get rid of these undesirable singularities, since in these cases our current understanding of physics breaks down and the predictability of theory is lost.
However, under a set of reasonable assumptions, Hawking and Penrose demonstrated that singularities will inevitably occur within the framework of GR~\cite{Penrose:1964wq,Hawking:1970zqf}.
Despite this, people can still circumvent singularities by relaxing some conditions in the singularity theorem. Since the discovery of the first regular black hole by Bardeen~\cite{Bardeen}, researchers have been exploring various models of such regular black holes~\cite{Dymnikova:1992ux,Simpson:2019mud,Simpson:2021dyo,Lan:2023cvz,Bambi:2023try,Modesto:2008im,Hayward:2005gi,Ghosh:2014pba,Frolov:2014jva,Frolov:2017rjz,Spallucci:2008ez,Koch:2014cqa,Bodendorfer:2019jay,Bodendorfer:2019nvy,Brahma:2020eos,Nicolini:2005vd,Bonanno:2000ep,Gambini:2008dy,Modesto:2004xx,Nicolini:2008aj}.
For example, a vacuum non-singular black hole is proposed, which for large radial coordinate $r$ coincides with the Schwarzschild solution and for small $r$ behaves like the de Sitter solution~\cite{Dymnikova:1992ux}. This is one of the so-called regular black holes with de Sitter core.
Similarly, one can also construct regular black holes with Minkowski core~\cite{Simpson:2019mud,Simpson:2021dyo}.
Another approach is to consider the quantum effect.
Some people believe that the emergence of the singularities implies that GR is incomplete at short distance, i.e., GR is an effective field theory.
It is natural to imagine that quantum effect will dominate in short distance, therefore a full quantum gravity theory may exclude singularities.
Loop Quantum Gravity (LQG) is one of the most popular quantum gravity theories which aims to merge quantum mechanics and GR.
Within the framework of LQG, the so-called loop black hole~\cite{Modesto:2008im} is projected by a simple modification of the holonomic Hamiltonian constraint.
It has many similarities with the Reissner-Nordstr\"{o}m (RN) black hole but without curvature singularities.
In particular, the region around $r = 0$ corresponds to another asymptotically flat region.
Other regular black holes by quantum gravity can be found in~\cite{Hayward:2005gi,Nicolini:2008aj,Nicolini:2005vd,Bonanno:2000ep,Gambini:2008dy,Modesto:2004xx}.

Recently, Lewandowski, Ma, Yang, and Zhang have proposed a new quantum black hole model in LQG~\cite{Lewandowski:2022zce}.
The spacetime of this model is a suitably deformed Schwarzschild black hole.
It is exterior of the quantum Oppenheimer-Snyder and Swiss Cheese models which projected by gravitational collapse and avoiding the singularity.
This quantum-corrected black hole exhibits the same asymptotic behavior as the Schwarzschild black hole and is stable against test scalar and vector fields by the analysis of quasinormal modes (QNMs)~\cite{Yang:2022btw}.
But it has several observable properties that differ from those of the Schwarzschild black hole, including its shadow and photon sphere~\cite{Zhang:2023okw}.
In particular, photons can enter the horizon of a companion black hole in the previous universe, travel through a highly quantum region, and then be captured by observers in our universe, resulting some photon spheres occur distinctly within the shadow region.
The images of supermassive compact objects recently photographed by the Event Horizon Telescope~\cite{EventHorizonTelescope:2019dse} provide a wealth of information about the surroundings of black holes and make it possible to examine the predictions of LQG by analysing these distinct photon spheres.

This quantum-corrected black hole has an event horizon and a Cauchy horizon as a price for avoiding the spacelike singularities just like many other quantum regular black holes and classical black holes, and the structure of photon spheres highly depends on these horizons.
However, the Cauchy horizon of classical RN black hole and Kerr black hole, as well as many other quantum regular black holes, is unstable~\cite{Mcnamara:1977,Mcnamara:1978,1982RSPSA.384..301C,Simpson:1973ua,Brown:2011tv,Iofa:2022dnc,Carballo-Rubio:2021bpr,Kazakov:1993ha}, which is consistent with the strong cosmic censorship hypothesis~\cite{R. Penrose:1974}.
This raises the question of the stability of the quantum-corrected black hole's Cauchy horizon and how this affects the behavior of the photon spheres.

For RN black holes, considering a perturbation originating from the event horizon, one can imagine that the blueshift effect during propagation could result in the divergence of the amplitude of the perturbation on the Cauchy horizon.
Since the metric of the quantum-corrected black hole is quite similar to that of RN black holes, it is natural to guess that, based on the same argument, the Cauchy horizon of the quantum-corrected black hole is also unstable.
However, for the loop black hole described in~\cite{Modesto:2008im}, the situation is significantly different from that of RN black holes~\cite{Brown:2010csa}.
Thus one should be careful when dealing with these quantum black holes.
We studied the flux and energy density of the test scalar field using Chandrasekhar's method thereby demonstrated the instability of the Cauchy horizon of quantum-corrected black holes. Therefore, this traditional physical argument for RN black hole remains valid for quantum-corrected black holes.
However, this approach does not take into account the backreaction of the spacetime and can not predict the fate of the Cauchy horizon.
To overcome this limitation, we then employ the generalized Dray-'t Hooft-Redmond (DTR) relation which is independent of Einstein's field equations.
Both methods show that the inner horizon is unstable and the generalized DTR relation further indicates that a null singularity will emerge.
Therefore this quantum Oppenheimer-Snyder and Swiss Cheese models can not be viewed as a regular black hole.
Thus one needs to further investigate about whether there is photon sphere that differs significantly from the Schwarzschild black hole's photon sphere.

This paper is organized as follows.
In section \ref{S2}, we briefly introduced the quantum-corrected black hole.
Then we use the method proposed by Chandrasekhar to study the flux and energy density of the test scalar field measured by free-falling observers.
In section \ref{S3}, we study the stability of the inner horizon by the generalized DTR relation.
When the ingoing shell approaches the inner horizon, we found that the mass inflation occurs and the Ricci scalar and Kretschmann scalar are both divergent.
Finally, we give conclusions and discussions in section \ref{S4}.

\section{The quantity measured by the free-falling observer}\label{S2}
The quantum Oppenheimer-Snyder and Swiss Cheese models are obtained in the framework of LQG~\cite{Lewandowski:2022zce}.
The spacetime of this model is connected by two regions with certain connection conditions. The interior is an Ashtekar-Pawlowski-Singh (APS)~\cite{Ashtekar:2006rx} dust ball and the exterior is the quantum-corrected black hole we are considering.
The quantum-corrected black hole obtained as a solution to the effective equations in certain LQG spherically symmetric model is a quantum extension of the Kruskal spacetime.
Its metric has a simple form
\begin{equation}
\label{metric}
\mathrm{d}s^2=-f(r)\mathrm{d}t^2+\frac{1}{f(r)}\mathrm{d}r^2+r^2(\mathrm{d}\theta^2+\sin^2\theta\mathrm{d}\phi^2 )\, ,
\end{equation}
with
\begin{equation}
f(r)=1-\frac{2m}{r}+\frac{\alpha m^2}{r^4}    \,,
\end{equation}
where $$\alpha=16\sqrt{3}\pi \gamma^3 \ell_p^2\, ,$$
$\ell_p$ is the Planck length and $\gamma$ denotes the Barbero-Immirzi parameter. The symbol $m$ is the mass parameter.
This metric coincides with that derived in \cite{Kelly:2020uwj}.
When $m>4\sqrt{3\alpha}/9$, it has two horizons.
The radius of the inner horizon
$r_-\approx \sqrt[3]{\alpha m/2} $
is nearly the Planck scale, while the radius of the outer horizon is approximately $2m$.
The surface gravity on the inner horizon and outer horizon are approximately
\begin{eqnarray}
\label{kappa}
\kappa_-&\approx& \sqrt[3]{\frac{m}{2\alpha^2}}     \,,\nonumber\\
\kappa_+&\approx& \frac{1}{4m}  \,.
\end{eqnarray}
Here and below we only consider the case that $m\gg4\sqrt{3\alpha}/9$, which might be interesting in astronomical observation.

In principle, to study the stability of the Cauchy horizon, we should consider the underline perturbation equation derived from the field equation and some fixed background. However, since we do not know the full effective gravitational field equation of LQG,  the necessary perturbation equation is absent.
Therefore we use the test scalar field equation $\Box \psi =0$, which can also reveal part of the nature of the perturbation equation. After seperating variables $\psi=\varphi e^{i\omega t}$, we get the radial wave equation
\begin{equation}
\frac{\mathrm{d}^2\varphi(r_*)}{\mathrm{d}r_*^2}+\left( \omega^2-V \right)\varphi(r_*)=0    \, ,
\end{equation}
where the potential function is
\begin{equation}
V(r)=f(r)\Big[\frac{l(l+1)}{r^2}+\frac{f^{\prime}(r)}{r}\Big]\, ,
\end{equation}
and we define the tortoise coordinate $r_*$ as
\begin{equation}
\label{r*}
\mathrm{d}r_*=\frac{1}{f}\mathrm{d}r\, .
\end{equation}
The detailed form of $r_*$ in terms of $r$ can be found in Appendix.
To study the stability of inner horizon, it is convenient to apply null coordinates. The advanced time is defined with the same convention as~\cite{1982RSPSA.384..301C}, i.e.,
\begin{equation}
v=r_*-t       \, .
\end{equation}
The limit $v\rightarrow+\infty$ corresponds to the Cauchy horizon of the black hole.
As $v$ tends to infinity, $t$ tends to
minus infinity, and $r$ approaches  $r_-$, i.e., the radius of the Cauchy horizon.
Thus, from the Eq.(\ref{r*}), we have
\begin{eqnarray}
r_*&\sim& -\frac{1}{2\kappa_-}\ln(r-r_-)\rightarrow +\infty  \, ,
\end{eqnarray}
and for finite $u=r_*+t$,
\begin{eqnarray}
v&=&r_*-t\sim2r_*\sim -\frac{1}{\kappa_-}\ln(r-r_-)   \, .
\end{eqnarray}
From the above results, we obtain
\begin{equation}
r-r_-\sim e^{-\kappa_-v}  \, ,  \qquad v\rightarrow+\infty  \, .
\end{equation}
Since  the black hole we considered is far from extreme, $f(r)$ is proportional to $(r-r_-)$ when $r\rightarrow r_-$.
So, near the Cauchy horizon, $f$ and $V$ have the forms
\begin{eqnarray}
\label{f}
f&\sim& e^{-\kappa_-v}    \, ,\qquad v\rightarrow+\infty       \, ,
\end{eqnarray}
and
\begin{eqnarray}
V&\sim& f \sim e^{-\kappa_-v}    \, ,\qquad v\rightarrow+\infty             \, ,
\end{eqnarray}
respectively. The above asymptotic behaviors are similar to the case of RN black hole.
Therefore, following the process by Chandrasekhar ~\cite{1982RSPSA.384..301C,1983mtbh.book.....C}, the flux measured by a free-falling observer is proportional to the square of the amplitude
\begin{equation}\label{amplitude}
\mathscr{F}=U^\alpha \partial_\alpha\psi   \, ,
\end{equation}
where $U^\alpha$ is the 4-velocity of the observer.
When $v\rightarrow+\infty$, which means the observer approaches the Cauchy horizon, the asymptotic behavior of the flux is established by
\begin{equation}\label{flux}
\mathscr{F}\rightarrow \text{Constant} \times e^{(\kappa_--\kappa_+)v}   \, .
\end{equation}
 One can find more details in the Appendix \ref{S5}.

From the Eq.(\ref{kappa}), we know $\kappa_-\gg \kappa_+>0$.
Since $\kappa_{-}$ is satisfied with
\begin{equation}
\kappa_-\approx \frac{1}{8\gamma^2}\sqrt[3]{\frac{m}{3\pi^2l_p^4}} \, ,
\end{equation}
where we have substituted $\alpha$ into Eq.(\ref{kappa}), the exponential part of Eq.(\ref{flux})  divergences seriously. Thus, the flux of radiation received by the geodesic observer is always divergent as he/she crosses the Cauchy horizon.
For the black hole with several solar  masses, it is not hard to find $\kappa_-\approx (10^{61}/\gamma^2) m/s^2$.
It is important to note that the value of $\kappa_-$ is consistently large determined by the inherent characteristics of the quantum-corrected black hole in LQG, in contrast to the case of RN black holes, where the value of $\kappa_-(>\kappa_+)$ can be set arbitrarily by adjusting the charge parameter.

However, not all of the black holes in LQG have divergent flux received by observers crossing the Cauchy horizon.
For the loop black hole with a macroscopic mass, the value of $\kappa_-$ is nearly zero resulting in observers receiving a finite flux and energy density~\cite{Brown:2011tv,Brown:2010csa}.
The Cauchy horizon is unstable when the mass is extremely small. In other words, the stability of the Cauchy horizon of the loop black hole depends on its mass.
In this sense, the loop black hole is more stable than classical black holes under this kind of scalar perturbation.
This is quite different from the quantum-corrected black hole which we are considering.
For the quantum-corrected black holes, regardless of their mass, the flux received by the observer is divergent. This is due to the fact that $\kappa_--\kappa_+$ is always positive (details can be found in Appendix \ref{S6}).
This  behavior is similar to that observed in the RN black hole.

In spite of this, if one considers a null dust shell with the stress-energy tensor
\begin{equation}
T_{\alpha\beta}=\mu(r,v)(\partial_\alpha v)( \partial_\beta v) \,,
\end{equation}
where
\begin{equation}
\mu=\frac{r^2}{4\pi r^4+a_0^2}\mathcal{L}(v)  \, ,
\end{equation}
and $a_0$ is the minimal area of the black hole, and $\mathcal{L}$ can be chosen as the one satisfying the Price law~\cite{Brown:2011tv}.
Then the energy density measured by observers crossing the Cauchy horizon is
\begin{eqnarray}
    \label{rhodust}
\rho_{\text{dust}}\sim e^{2\kappa_- v}v^{-p}   \, ,
\end{eqnarray}
which is divergent.
Consequently, the Cauchy horizon of the loop black hole is unstable. For the quantum-corrected black hole, by the same logic, we can get the same result as the one in Eq.(\ref{rhodust}). So the Cauchy horizon of the quantum-corrected black hole is also unstable in the model of the null dust shell.

Actually, since we are considering the test scalar field, it is natural to adopt the energy density  of the stress-energy tensor (measured by free-falling observers)
\begin{equation}\label{T}
T_{\alpha\beta}=\frac{1}{2}\Big(\nabla_\alpha \psi\nabla_\beta\psi-\frac{1}{2}g_{\alpha\beta}\nabla_\gamma \nabla^\gamma\psi\Big)   \,.
\end{equation}
The calculation in Appendix \ref{S5} shows the energy density measured by the observer is given by
\begin{equation}
\rho_\psi=T_{\alpha\beta}U^\alpha U^\beta\sim e^{2(\kappa_- -\kappa_+) v}\rightarrow+\infty  \,, \quad v\rightarrow+\infty     \,.
\end{equation}
As a result, for the loop black hole~\cite{Brown:2011tv} with macroscopic mass, the flux of the scalar field and its energy density are finite, while the energy density of the null dust is divergent.
In contrast, these quantities are all divergent for the quantum-corrected black hole (\ref{metric}).

In this section, we even treated matter fields classically near the Cauchy horizon, whose radius is close to the Planck scale.
In fact, these classical matter fields are substitutes for more complicated quantum matter fields.
We will give a brief discussion on the validity of our calculation in section \ref{S4}.

\section{Mass inflation}  \label{S3}
In many classical black holes, such as the RN black hole, the mass function grows unbounded due to the blueshift of the ingoing flux.
This phenomenon is known as mass inflation~\cite{Ori:1991zz,Poisson:1990eh}, resulting in the emergence of a null singularity rather than a timelike singularity.
The ingoing flux can be generated when the outgoing flux is reflected by the potential.
We have demonstrated its divergent near the Cauchy horizon.
Therefore, one should calculate both the ingoing and outgoing flux when considering the perturbation.
In fact, Poisson and Israel~\cite{Poisson:1990eh} found that the presence of an ingoing flux alone is insufficient to cause mass inflation in the RN black hole. Mass inflation only occurs when both ingoing and outgoing fluxes are present.

Poisson and Israel uesd the null dust shell model to investigate mass inflation of RN black holes, and derived the DTR relation from the Einstein's equations~\cite{Poisson:1990eh}.
Although the DTR relation is a powerful tool to study mass inflation, this approach encounters difficulties when applied to a variety of quantum black holes due to the lack of a field equation.
Fortunately, a generalized DTR relation that does not rely on the field equations was developed by Eric and Robert~\cite{Brown:2011tv} to study mass inflation.
They used a model in which the ingoing and outgoing fluxes are represented by two null shells, dividing spacetime into four parts.
The metric of each part is characterized by its mass function.
In case of RN black hole, mass inflation occurs when the ingoing flux approaches the Cauchy horizon.

For classical black holes, for example, RN black hole, mass inflation usually implies instability of the Cauchy horizon and appearance of null singularity.
But we do not know whether the instability of the Cauchy horizon is always accompanied by the mass inflation.
For quantum black holes, things may be different.
For example, the Cauchy horizon of the loop black hole~\cite{Brown:2011tv} is unstable because of the divergent energy density and Ricci scalar, while its mass function is convergent.
In contrast, when its mass parameter is infinity, the Ricci scalar and  Kretschmann scalar both remains finite~\cite{Brown:2011tv}.
Therefore, it is worth to study the stability and mass inflation of the quantum-corrected black hole.
In this section, we use the generalized DTR relation to study the stability of quantum corrected black holes and further investigate whether their internal structure will affect shadow images.

The radiation inside the black hole is replaced by two null shells as shown in the Penrose diagram Fig.1.
\begin{figure}[htb]
\label{penrose}
  \centering
  \includegraphics[width=8cm]{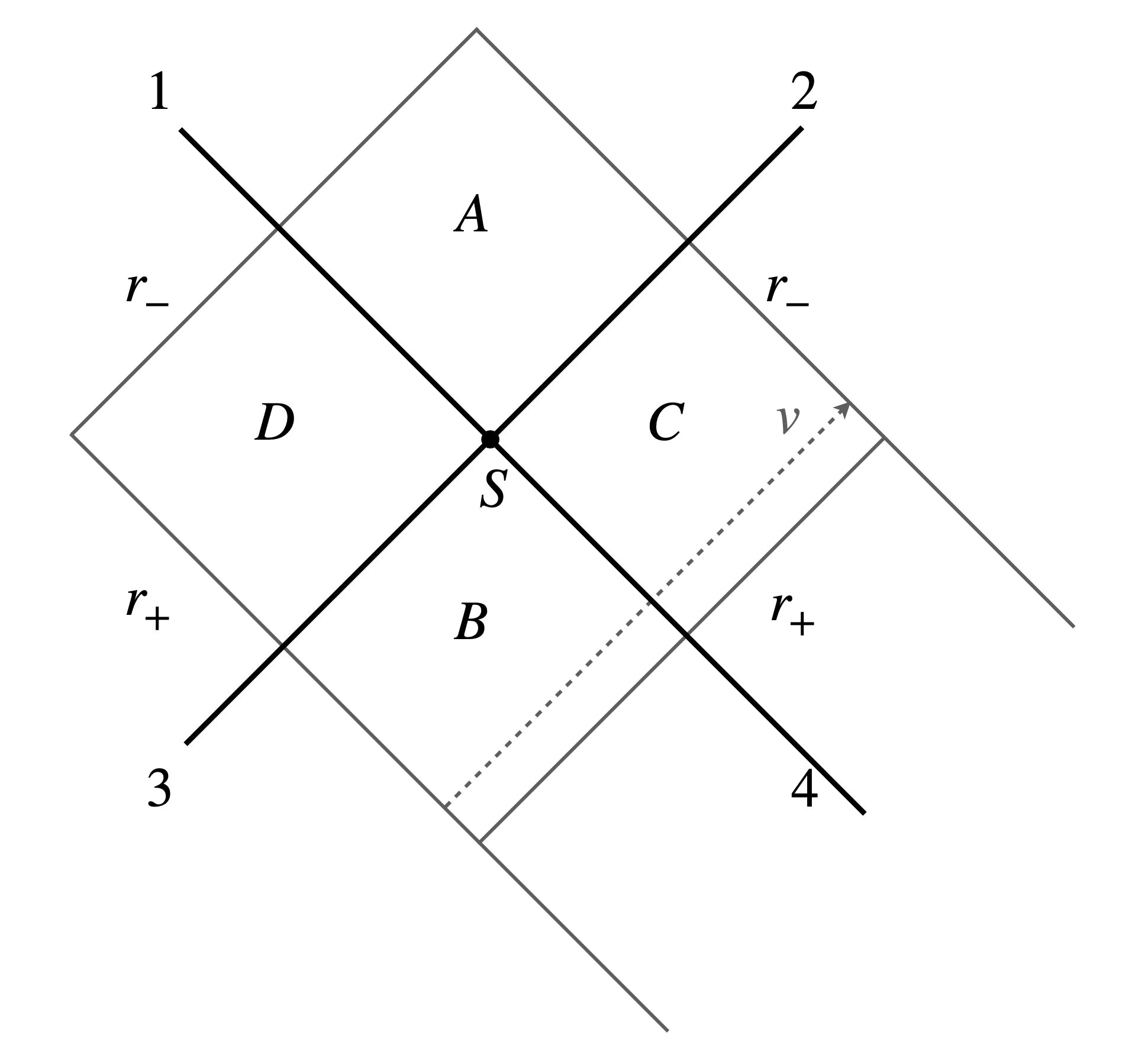}
  \caption{The Penrose diagram between two horizons of the quantum-corrected black hole. 4-1 is the ingoing shell, 3-2 is the outgoing shell,  and they cross each other at the two-dimensional spherical shell $S$ and divide the spacetime between the two horizons into four parts. The coordinate $v$ is shown as the dotted line.}
\end{figure}
The line 4-1 represents the ingoing shell and the line 3-2 represents the outgoing one.
They cross each other at $S$, where $S$ is a two-dimensional spherical shell and the coordinate $r$ is continuous on $S$.
The $v$ coordinate line is parallel to the outgoing shell 3-2, and thus the ingoing shell 4-1 is labeled by $v$.
The two shells divide the spacetime between outer horizon and inner horizon  into four regions, $A$, $B$, $C$, and $D$, respectively.
Each region is characterized by its mass function.
For example, the mass $m_A$ remains constant throughout the region $A$.
The mass of the ingoing shell equals $m_C-m_B$ and is assumed to be the Price influx $v^{-p}$, which is the classical result relying on Einstein's field equations ~\cite{Price:1972pw} and $p\geqslant12$, i.e.,
\begin{equation}
\label{mBmC}
m_C-m_B\sim v^{-p}  \,.
\end{equation}
The mass of outgoing shell is $m_D-m_B$ and is assumed not vanishing.

If we change the ``position" of the ingoing shell, then the mass function will be changed too.
This means $m_i, i=A,B,C,D$  are the functions of $v$.
Different $v$ represents the different position of ingoing shell and the different spacetime.
In the following, we will  demonstrate that when the ingoing shell approaches the inner horizon, i.e. $v\rightarrow\infty$, the mass inflation occurs.

Following the calculation in~\cite{Brown:2011tv},
the intersection $S$ divides two shells into four parts which we refer to as $1S$, $2S$, $3S$ and $4S$.
The normal vector of each part is
\begin{equation}
\label{li}
\ell^\alpha_i=\frac{\partial x^\alpha}{\partial \lambda_i}   \,,
\end{equation}
where $\lambda_i$ is the affine parameter of the null geodesic of the shell and $i=1, 2, 3, 4$ denotes the shell $iS$.
We also define two frames on the shell $e^\alpha_a$, $a=\theta, \phi$.
Since the shell is null, we have (the index $i$ is omitted)
\begin{equation}
\ell^\alpha \ell_\alpha=e_a^\alpha \ell_\alpha=0   \,.
\end{equation}
Thus, the induced metric on the shell is
\begin{equation}
\sigma_{ab}=g_{\alpha\beta}e_a^\alpha e_b^\beta  \,.
\end{equation}
Conversely, with the help of two null vectors, one can represent $g^{\alpha\beta}$ using $\sigma^{ab}$.
Take region $A$ as an example,
\begin{equation}
\label{g}
g_{(A)}^{\alpha\beta}=2\frac{\ell_1^{(\alpha}\ell_2^{\beta)}}{\ell_1\cdot \ell_2}+\sigma^{ab}e_a^\alpha e_b^\beta  \,.
\end{equation}
For convenience, the two frames $e^\alpha_a$ can be chosen by $e^\alpha_\theta=(0,0,1,0)$ and $e^\alpha_\phi=(0,0,0,1)$.
The induced metric is given by
\begin{eqnarray}
\sigma_{a b}=r^2\left[\begin{array}{cc}
1 & 0 \\
0 & \sin ^2 \theta
\end{array}\right]   \,.
\end{eqnarray}
Since the entries of the induced metric are like scalars with respect to the full four dimensional spacetime, the extrinsic curvature of the shell is
\begin{equation}
K_{a b} \equiv \frac{1}{2} \mathscr{L}_{\ell}\sigma_{a b}=\frac{1}{2} \ell^\alpha\left(\sigma_{a b}\right)_{, \alpha}  \,,
\end{equation}
and its trace is
\begin{eqnarray}
\label{K}
K&\equiv&\sigma^{ab}K_{ab}        \,\nonumber\\
&=&\frac{1}{2} \ell^\alpha \sigma^{ab}\left(\sigma_{a b}\right)_{, \alpha}        \,\nonumber\\
&=&\frac{1}{2} \ell^\alpha \frac{1}{\sigma}\partial_\alpha\sigma        \,\nonumber\\
&=&\frac{1}{2r^4} \ell^\alpha \partial_\alpha r^4    \,,
\end{eqnarray}
where $\sigma$ is the determinant of $\sigma_{ab}$.

Since the intersection $S$ is a two-dimensional spherical shell, there are only two linearly independent null vectors normal to it.
However, from (\ref{li}), we obtained a total of 4 vectors, $\ell^\alpha_1, \ell^\alpha_2, \ell^\alpha_3$ and $\ell^\alpha_4$ that are normal to $S$.
Therefore, there are only two linearly independent $\ell^\alpha_i$ on $S$.
We choose $\ell_1$ and $\ell_2$ to be the independent null normal vectors,
\begin{equation}
\ell_1=a\ell_4 \,,  \qquad    \ell_2=b\ell_3    \,,
\end{equation}
where $a$ and $b$ are some coefficients.
Then we can get an identity on $S$,
\begin{equation}
\label{l=l}
(\ell_1\cdot \ell_2)(\ell_3\cdot \ell_4)=(\ell_2\cdot \ell_4)(\ell_1\cdot \ell_3)  \,.
\end{equation}
We define four functions on $S$, namely, $Z_A, Z_B, Z_C$ and $Z_D$,
\begin{equation}
\label{Z}
Z_A \equiv \frac{K_1 K_2}{\ell_1 \cdot \ell_2}   \,,
\end{equation}
and similar definitions holds for $B, C$ and $D$, where $K_i$ is the trace of extrinsic curvature of the shell $iS$ and the subscripts of $Z$ corresponds to the spacetime regions shown in Fig.1.
After that, (\ref{l=l}) can be extended to
\begin{equation}
\label{Z=Z}
|Z_A|\cdot |Z_B|=|Z_C|\cdot|Z_D|    \qquad \text{on} \,\, S  \,.
\end{equation}
By taking the absolute value of $Z$'s, we avoid ambiguity of the signatures of $\ell_i$ since a reparameterization $\lambda_i\rightarrow -\lambda_i$ would change it.
It is worth noting that this relation is independent of the field equation.
From (\ref{K}) and (\ref{Z}) we obtain
\begin{equation}
Z_A= \frac{1}{4r^8}\frac{\ell_1^\alpha \ell_2^\beta (\partial_\alpha r^4)(\partial_\beta r^4)}{\ell_1\cdot \ell_2}  \,.
\end{equation}
The expression for $Z_A$ is not explicit up to now.
To relate $Z_A$ to the metric $g^{\alpha\beta}_{(A)}$ in region $A$, from (\ref{g}), we get
\begin{equation}
g^{\alpha\beta}_{(A)}(\partial_\alpha r^4)(\partial_\beta r^4)=2\frac{\ell_1^\alpha \ell_2^\beta (\partial_\alpha r^4)(\partial_\beta r^4)}{\ell_1\cdot \ell_2}  \,,
\end{equation}
where the term $e^\alpha_a\partial_\alpha r^4$ is zero because $e^\alpha_\theta\partial_\alpha=\partial_\theta$ and the similar case for $e^\alpha_\phi$ holds.
Hence $Z_A$ can be expressed in the form of $g^{\alpha\beta}_{(A)}$,
\begin{equation}
Z_A=\frac{1}{8r^8}g^{\alpha\beta}_{(A)}(\partial_\alpha r^4)(\partial_\beta r^4)=\frac{2}{r^2}g^{rr}_{(A)}=\frac{2}{r^2}f_A  \,,
\end{equation}
and Eq.(\ref{Z=Z}) can be reduced to
\begin{equation}
|f_A|\cdot|f_B|=|f_C|\cdot|f_D|      \,,
\end{equation}
or
\begin{equation}
\label{fA}
|f_A|=\left|\frac{f_C}{f_B}\right|\cdot|f_D|  \,.
\end{equation}

When $v\rightarrow\infty$, the ingoing shell almost overlaps with the inner horizon.
Since region $B$ is connected to the asymptotic flat region located outside of the event horizon, both regions share the same mass parameter $m_B$.
Therefore, the mass of the black hole detected by an external observer is $m_B$.
This implies that the inner horizon locates at $r_-$, which satisfies
\begin{equation}
f_B(r_-)=1-\frac{2m_B}{r_-}+\frac{\alpha m_B^2}{r_-^4}=0   \,.
\end{equation}

When the ingoing shell approaches the inner horizon, i.e. $v\rightarrow+\infty$ and $r\rightarrow r_-$, the Eqs.(\ref{mBmC}) and (\ref{f}) show that $m_C \approx m_B$ and $f_B\approx0$, which leads to $f_C \approx 0$ too.
In contrast, $f_D$ is a non-vanishing quantity of finite magnitude since $m_D\neq m_B$.
These make it hard to determine the right hand side of (\ref{fA}).
Using $m_C=m_B+v^{-p}$, when $v\rightarrow+\infty$, we have
\begin{eqnarray}
f_C&=&1-\frac{2m_C}{r}+\frac{\alpha m_C^2}{r^4}+\mathcal{O}\left( v^{-2p} \right)       \,\nonumber\\
&=&1-\frac{2m_B}{r}-\frac{2v^{-p}}{r}+\frac{\alpha}{r^4}\left(m_B^2+2m_Bv^{-p}\right)+\mathcal{O}\left( v^{-2p} \right)       \,\nonumber\\
&=&f_B-\frac{2}{r}v^{-p}+\frac{2\alpha}{r^4}m_Bv^{-p} +\mathcal{O}\left( v^{-2p} \right)      \,\nonumber\\
&=&f_B+2v^{-p}\left(\frac{\alpha m_B}{r^4}-\frac{1}{r}\right)+\mathcal{O}\left( v^{-2p} \right)       \,.
\end{eqnarray}
The second term of the last line is finite as $r\rightarrow r_-$.
As a result,
\begin{equation}
\frac{f_C}{f_B}\sim  f_B^{-1}v^{-p} \sim v^{-p}e^{\kappa_-v}\rightarrow+\infty \,, \qquad v\rightarrow+\infty\,.
\end{equation}
Therefore, from Eq.(\ref{fA}),
\begin{equation}
f_A=1-\frac{2m_A}{r_-}+\frac{\alpha m_A^2}{r_-^4}    \,,
\end{equation}
is divergent.
Because $\alpha$ is as the Planck scale and $r_-$ is the root of $f_B=0$ rather than $f_A=0$, the only way for $f_A$ to be divergent is that
\begin{equation}
m_A\rightarrow+\infty  \,.
\end{equation}
This means the mass inflation always occurs.
The situation is very similar to the RN black hole, but is different from the loop black hole.

The Ricci scalar and the Kretschmann scalar of region $A$ as $v\rightarrow\infty$ are
\begin{eqnarray}
R&=& -\frac{6\alpha m_A^2}{r_-^6} \rightarrow -\infty \, ,
\end{eqnarray}
and
\begin{eqnarray}
R_{\alpha\beta\gamma\delta}R^{\alpha\beta\gamma\delta}&=& \frac{468\alpha^2m_A^4}{r_-^{12}}-\frac{240\alpha m_A^3}{r_-^9}+\frac{48m_A^2}{r_-^6} \rightarrow+\infty \,.
\end{eqnarray}
That is, the phenomenon of the scalar curvature singularity happens at the inner horizon.
A null singularity of the quantum-corrected black hole will appear.

\section{conclusions and discussion}\label{S4}

In this paper, we have analysed the stability of the inner horizon of the quantum-corrected black hole by two ways.
First, we calculate the flux or energy density of the test scalar field following the process by Chandrasekhar.
They are both divergent as the observer approaching to the Cauchy horizon.
Second, the generalized DTR relation allows us to study the mass inflation of the inner horizon without the field equation.
As a result, the mass inflation occurs and the Ricci scalar and the Kretschmann scalar are both divergent near the inner horizon.
This implies that the inner horizon is unstable and will become a null singularity.

In~\cite{Zhang:2023okw}, the authors postulated that certain photons could traverse the quantum region within the inner horizon, subsequently being captured by an observer, thereby generating a novel image.
This process, however, relies on the internal structure of the black hole, particularly the stability of the inner horizon.
Our analysis suggests that photons are unable to pass through the inner horizon and escape from a white hole due to the instability of the inner horizon, which eventually transforms into a null singularity.
Despite this, they utilized a recently proposed spacetime structure~\cite{Han:2023wxg}, which was developed by ``cutting and stitching'' the original spacetime and includes a specific region denoted as $\mathcal{B}$.
This new spacetime model describes the transformation of a black hole into a white hole located in the same asymptotic region.
The causal structure of this new spacetime model differs from the original one, and the complicated $\mathcal{B}$ region makes it difficult to determine whether the coordinate $v$, whose ability to tends to infinity plays a crucial role in our calculation, can indeed tends to infinity.
This makes the presence of such novel image possible.
However, this $\mathcal{B}$ region is supposed to be a highly ``quantum" region and be of Planckian size.
The coordinate $v$ can still attain a large value, and the free-falling observer can still receive a significant energy flux.
Consequently, it remains unclear whether this novel phenomenon will still occur.
Therefore, it may be challenging to extract information about quantum gravity through this approach.
These require further investigation.

To circumvent the singularity within black holes, regular black holes are projected in several approaches to quantum gravity.
Although these regular black holes have nonsingular cores, some of them have more than one horizon.
The quantum Oppenheimer-Snyder and Swiss Cheese models considered in~\cite{Lewandowski:2022zce} are consists of Ashtekar-Pawlowski-Singh (APS) model and the quantum-corrected black hole.
The spacetime has no singularity and its exterior approximates Schwarzschild spacetime.
Based on our analysis, however, the inner horizon of the quantum-corrected black hole is unstable.
The inner horizon will become null singularity, which takes an indication that this model may not be the definitive endpoint of a quantum black hole without singularity.

Physics theory breaks down at the singularity which can affect the causal structure of spacetime.
Thus, the weak cosmic censorship conjecture (WCCC) is proposed that singularities are hidden behind the event horizon~\cite{Penrose:1969pc}.
Further more, the strong cosmic censorship conjecture (SCCC) states that any observer, even those inside the event horizon, can receive no information from a singularity.
In other words, SCCC posits the existence of a Cauchy surface within spacetime, rendering it predictable.
When there exist Cauchy horizon, the predictability of the theory is lost and SCCC breaks down\cite{R. Penrose:1974}.
It seems that the quantum effect would violate the SCCC due to the presence of Cauchy horizon.
Fortunately, our result shows that the Cauchy horizon of the quantum-corrected black hole is unstable, therefore prevents the extension of the spacetime beyond the Cauchy horizon and preserving the SCCC.

In~\cite{Carballo-Rubio:2019fnb}, the authors had proven all regular black holes have inner horizons.
Some famous examples are the loop black hole~\cite{Modesto:2008im}, the Hayward black hole~\cite{Hayward:2005gi}, the quantum-corrected black hole~\cite{Lewandowski:2022zce} and so on~\cite{Modesto:2008im,Hayward:2005gi,Spallucci:2008ez,Bonanno:2000ep,Gambini:2008dy,Koch:2014cqa,Bodendorfer:2019jay,Bodendorfer:2019nvy,Brahma:2020eos,Nicolini:2005vd,Ghosh:2014pba,Frolov:2014jva,Frolov:2017rjz}.
The inner horizons of the black holes metioned above are unstable and will evolve into a null singularity~\cite{Brown:2011tv,Iofa:2022dnc,Carballo-Rubio:2021bpr}.
Although these black holes were proposed to eliminate singularities, the instability of their inner horizons eventually produces null singularity.
On the other hand, a regular black hole without mass inflation is proposed and it has a stable inner horizon~\cite{Carballo-Rubio:2022kad}.
However, the precise physical process that leads to the formation of such black hole remains a mystery.
Thus the regular black hole models requires further investigation.
Meanwhile, it remains an open question as to whether the Cauchy horizons of other regular black holes are stable.
The methods used in this article can be employed to study this topic, and we will conduct further research.

Finally, we offer some discussion on the validity of our calculation.
Our calculations are based on the assumptions that the matter fields, e.g., the test scalar field, can be treated classically.
However, the inner horizon, which plays a crucial role in our calculations, has an extremely small radius that even approaches the Planck scale.
So one may wonder whether this assumption is reasonable.
In fact, not only the classical matter fields are invalid at this scale, but classical geometry is also not applicable.
As a consequence, the spacetime model and classical observers do not exist in principle.
Nevertheless, the quantum-corrected metric~(\ref{metric}) serves as an effective solution that encodes quantum effects in a classical manner.
This effective model allows us to represent complicated quantum matter fields using more comprehensible classical matter fields.
Without this assumption, our understanding of spacetime, the inner horizon, and observers would not be possible.
Our calculations are based on this effective model and the results are only valid under this assumption.

\section*{Acknowledgement}

We would like to thank Li Li and Run-Qiu Yang for their useful discussions and comments.
We also thanks Cong Zhang for his valuable communication.
This work was supported in part by the National Natural Science Foundation of China with grants No.12075232 and No.12247103.
It is also supported by the Fundamental Research Funds for the Central Universities under Grant No.WK2030000036 and the National Key R\&D Program of China Grant No.2022YFC2204603.

\appendix
\section{}\label{S6}
In this appendix, we show that $\kappa_--\kappa_+$ is always positive for the quantum-corrected black hole.
Given $f(r_\pm)=0$, i.e.,
\begin{eqnarray}
1-\frac{2m}{r_+}+\frac{\alpha m^2}{r_+^4}&=&0   \,,\nonumber\\
1-\frac{2m}{r_-}+\frac{\alpha m^2}{r_-^4}&=&0    \,,
\end{eqnarray}
one obtains
\begin{eqnarray}
m&=&\frac{r_+^3+r_+^2 r_-+r_+ r_-^2+r_-^3}{2(r_+^2+r_+ r_-+r_-^2)}   \,,\nonumber\\
\alpha&=&\frac{4r_+^3 r_-^3(r_+^2+r_+ r_-+r_-^2)}{(r_+^3+r_+^2 r_-+r_+ r_-^2+r_-^3)^2}  \,.
\end{eqnarray}
Therefore, $f$ can be written as
\begin{equation}
f(r)=1-\frac{r_+^3+r_+^2 r_-+r_+ r_-^2+r_-^3}{r(r_+^2+r_+ r_-+r_-^2)}+\frac{r_+^3 r_-^3}{r^4(r_+^2+r_+ r_-+r_-^2)}  \,.
\end{equation}
The difference between the two surface gravity is
\begin{eqnarray}
\kappa_--\kappa_+&=&\frac{3r_+^5-r_+^4 r_--2r_+^3 r_-^2-2r_+^2 r_-^3-r_+ r_-^4+3r_-^5}{2r_+^2 r_-^2(r_+^2+r_+ r_-+r_-^2)}   \,\nonumber\\
&=&\frac{(r_+-r_-)^2(r_++r_-)(3r_+^2+2r_+ r_-+3r_-^2)}{2r_+^2 r_-^2(r_+^2+r_+ r_-+r_-^2)}   \,\nonumber\\
&>&0   \,.
\end{eqnarray}
This is similar to that of RN black holes.

\section{}\label{S5}
In this appendix, we follow the process of Chandrasekhar~\cite{1982RSPSA.384..301C,1983mtbh.book.....C} to calculate the flux and the energy density measured by a free-falling observer.
For the metric (\ref{metric}), $f$ can be factorized as
\begin{equation}
f(r)=\frac{1}{r^4}(r-r_+)(r-r_-)(r^2+pr+q)=\left(1-\frac{r_+}{r}\right)\left(1-\frac{r_-}{r}\right)\left(1+\frac{p}{r}+\frac{q}{r^2}\right)   \,,
\end{equation}
where $p$ and $q$ are constant about $\alpha$ and $m$.
Then the tortoise coordinate $r_*$ is
\begin{eqnarray}
r_*&=&\frac{r_-^4}{r_--r_+}\frac{1}{r_-^2+pr_-+q}\ln(r-r_-)-\frac{r_+^4}{r_--r_+}\frac{1}{r_+^2+pr_++q}\ln(r_+-r)      \nonumber\\
&&+r+C_1 \arctan\frac{2r+p}{\sqrt{-p^2+4q}}+C_2\ln(r^2+pr+q)      \,,
\end{eqnarray}
or
\begin{eqnarray}
r_*&=&\frac{1}{2\kappa_+}\ln(r_+-r)-\frac{1}{2\kappa_-}\ln(r-r_-)+r+C_1 \arctan\frac{2r+p}{\sqrt{-p^2+4q}}+C_2\ln(r^2+pr+q)   \,,
\end{eqnarray}
where $C_1, C_2$ are constants. When $p, q\rightarrow0$, $C_1, C_2\rightarrow0$.
Obviously $r_*\rightarrow+\infty$ as $r\rightarrow r_-$ and $r_*\rightarrow-\infty$ as $r\rightarrow r_+$.

The mode function $Z$ satisfies
\begin{equation}
\frac{\mathrm{d}^2Z}{\mathrm{d}x^2}+\left( \omega^2-V \right)Z=0    \,,
\end{equation}
with the potential having the asymptotic behavior
\begin{eqnarray}
V&\rightarrow& e^{-2\kappa_- x}   \,, \qquad x\rightarrow +\infty  \,,\nonumber\\
V&\rightarrow& e^{2\kappa_+ x}   \,,\qquad  x\rightarrow -\infty  \,.
\end{eqnarray}
The two solutions $f_1$ and $f_2$ have the asymptotic behavior
\begin{eqnarray}
&f_1\rightarrow e^{-i\omega x} \,, \qquad x\rightarrow+\infty   \,,\nonumber\\
&f_2\rightarrow  e^{i\omega x} \,,  \qquad x\rightarrow-\infty    \,.
\end{eqnarray}
Here, $r_*$ in section \ref{S2} is denoted as $x$, and tends to $v/2$ as $v\rightarrow+\infty$.

For the quantum-corrected black hole, the boundary conditions outside the event horizon are
\begin{eqnarray}
Z&\rightarrow& e^{i\omega x}+R e^{-i\omega x} \,,\,\qquad x\rightarrow +\infty    \,,\nonumber\\
Z&\rightarrow& T e^{i\omega x}  \,, \qquad\qquad\qquad x\rightarrow -\infty    \,,
\end{eqnarray}
where $R(\omega)$ and $T(\omega)$ are the reflection and transmission coefficients respectively.
Because $(f_1, f_1^*)$ and $(f_2, f_2^*)$ are pairs of independent solutions, it follows that there is an expansion
\begin{equation}
f_2(\omega)=\frac{1}{T(\omega)}f_1(-\omega)+\frac{R(\omega)}{T(\omega)} f_1(\omega)  \,.
\end{equation}
We can then deduce that
\begin{equation}
f_2(\omega)=\frac{1}{T(\omega)}f_1(-\omega)+\frac{R(\omega)}{T(\omega)} f_1(\omega)  \,.
\end{equation}
From this we can get
\begin{eqnarray}
\left[f_1(\omega,x),f_1(-\omega,x)\right]&=&-2i\omega                \,,\nonumber\\
\left[f_2(\omega,x),f_2(-\omega,x)\right]&=&2i\omega                  \,,\nonumber\\
\left[f_1(\omega,x),f_2(\omega,x)\right]&=&\frac{1}{T}(-2i\omega)           \,,\nonumber\\
\left[f_2(\omega,x),f_1(-\omega,x)\right]&=&\frac{R(\omega)}{T(\omega)}(-2i\omega)   \,,
\end{eqnarray}
where $[f,g]=fg'-f'g$ is the Wronskian of the functions $f$ and $g$.
Then we can deduce that
\begin{eqnarray}
\frac{1}{T(\omega)}&=&-\frac{1}{2i\omega}[f_1(\omega,x),f_2(\omega,x)]    \,,\nonumber\\
\frac{R(\omega)}{T(\omega)}&=&-\frac{1}{2i\omega}[f_2(\omega,x),f_1(-\omega,x)]  \,.
\end{eqnarray}
Between the outer horizon and inner horizon, the boundary conditions are
\begin{align}
&Z\rightarrow A(\sigma)e^{-i\sigma x}+B(\sigma) e^{i\sigma x} \,, \qquad x\rightarrow +\infty    \,,\nonumber\\
&Z\rightarrow  e^{-i\sigma x} \,,  \qquad\qquad\qquad\qquad\quad x\rightarrow -\infty    \,.
\end{align}
The coefficient $A$ and $B$ can be got from relation between the three boundary conditions above, and finally the solution between the two horizons is
\begin{align}
&Z\rightarrow \frac{1}{T(-\sigma)}e^{-i\sigma x}+\frac{R(-\sigma)}{T(-\sigma)} e^{i\sigma x}\,, \qquad x\rightarrow +\infty    \,,\nonumber\\
&Z\rightarrow  e^{-i\sigma x}\,,  \qquad \qquad \qquad \qquad \quad\qquad x\rightarrow -\infty    \,.
\end{align}
Using Eddington coordinates $u=r_*+t\, , v=r_*-t$, and multiplied by the time factor $e^{i\omega t}$, the time dependent solution is
\begin{align}
&Z\rightarrow \frac{1}{T(-\sigma)}e^{-i\sigma v}+\frac{R(-\sigma)}{T(-\sigma)} e^{i\sigma u}\,, \qquad x\rightarrow +\infty    \,,\nonumber\\
&Z\rightarrow  e^{-i\sigma v}\,,   \qquad \qquad \qquad \qquad\qquad\quad x\rightarrow -\infty    \,.
\end{align}
Finally, the solution of scalar field $\psi$ constructed from the mode functions $Z$ and the initial amplitude $\mathcal{L}$ is
\begin{equation}
\psi=\int_{-\infty}^{+\infty}\mathcal{L}(\omega)Z(\omega,v,u)d\omega   \,.
\end{equation}
Near the Cauchy horizon, it can be written as
\begin{equation}
\psi\equiv X(v)+Y(u)     \,,
\end{equation}
where
\begin{eqnarray}\label{XY}
X(v)&=&\int_{-\infty}^{+\infty}\mathcal{L}(\omega)[A(\omega)-1]e^{-i\omega v}d\omega   \,,\nonumber\\
Y(u)&=&\int_{-\infty}^{+\infty}\mathcal{L}(\omega)B(\omega)e^{i\omega u}d\omega         \,.
\end{eqnarray}
Thus
\begin{eqnarray}\label{X'Y'}
X_{,-v}(v)&=&\frac{\partial X}{\partial(-v)}
=\int_{-\infty}^{+\infty}i\omega\mathcal{L}(\omega)\left[\frac{1}{T(-\omega)}-1\right]e^{-i\omega v}d\omega     \,\nonumber\\
&=&\int_{-\infty}^{+\infty}\Big{\{}[f_1(-\omega,x),f_2(-\omega,x)]-2i\omega\Big{\}}\mathcal{L}(\omega)e^{-i\omega v}d\omega    \,,\nonumber\\
Y_{,u}(u)&=&\frac{\partial Y}{\partial u}
=\int_{-\infty}^{+\infty}i\omega\mathcal{L}(\omega)\frac{R(-\omega)}{T(-\omega)}e^{i\omega v}d\omega    \,\nonumber\\
&=&\int_{-\infty}^{+\infty}[f_2(-\omega,x),f_1(\omega,x)]\mathcal{L}(\omega)e^{i\omega u}d\omega    \,.
\end{eqnarray}
The components of the 4-velocity of the free-falling observer are
\begin{eqnarray}
u^t&=&\frac{E}{f}   \,,\nonumber\\
u^{r_*}&=&\frac{\sqrt{E^2-f}}{f}   \,,\nonumber\\
u^\theta&=&u^\phi=0             \,,
\end{eqnarray}
where $E$ is a constant, and therefore the amplitude (\ref{amplitude}) becomes
\begin{equation}\label{F}
\mathscr{F}=\frac{1}{f}E\psi_{,t}+\frac{1}{f}\sqrt{E^2-f}\psi_{,r_*}=\frac{1}{f}\left[ (E-\sqrt{E^2-f})X_{,-v}+(E+\sqrt{E^2-f})Y_{,u}  \right]   \,.
\end{equation}
The Cauchy horizon is located at $v\rightarrow+\infty$ or $u\rightarrow+\infty$.
When $v\rightarrow+\infty$, the first term of (\ref{F}) is divergent, and
\begin{equation}
\mathscr{F}\rightarrow-\frac{2|E|}{f}X_{,-v}=-\frac{r_-^4}{(r_--r_+)(r_-^2+sr_-+t)}\frac{1}{r-r_-}2|E|X_{,-v} \sim \mathscr{F}_v   \,,
\end{equation}
where
\begin{equation}
\mathscr{F}_v\equiv e^{\kappa_- v}X_{,-v} \,.
\end{equation}
Similarly, when $u\rightarrow+\infty$,
\begin{equation}
\mathscr{F}\sim \mathscr{F}_u\equiv e^{\kappa_- u}Y_{,u}   \,.
\end{equation}
From (\ref{X'Y'}), we obtain
\begin{eqnarray}
\mathscr{F}_v&=&e^{\kappa_- v}\int_{-\infty}^{+\infty}\Big{\{}[f_1(-\omega,x),f_2(-\omega,x)]-2i\omega\Big{\}}\mathcal{L}(\omega)e^{-i\omega v}d\omega    \,,\nonumber\\
\mathscr{F}_u&=&e^{\kappa_- u}\int_{-\infty}^{+\infty}[f_2(-\omega,x),f_1(\omega,x)]\mathcal{L}(\omega)e^{i\omega u}d\omega    \,.
\end{eqnarray}
Their behavior depend on poles of $f_1$ and $f_2$.
Analysing the analytic properties of $f_1$ and $f_2$, we find that the poles of $f_1(\omega,x)$ is $in\kappa_-$, where $n$ is a positive integer, and the poles of $f_2(-\omega,x)$ is $\omega=-in\kappa_+$, and $\mathcal{L}$ is analytic in the upper half complex $\omega$-plane.
A detailed proof can be found in~\cite{1982RSPSA.384..301C,1983mtbh.book.....C}.

The asymptotic behavior of $\mathscr{F}_u$ as $u\rightarrow+\infty$ is
\begin{equation}\label{Fu}
\mathscr{F}_u\sim\sum_n e^{\kappa_-u} e^{-n\kappa_-u} \,.
\end{equation}
It is convergent since $n\geqslant1$.
Therefore, the divergent term of $\mathscr{F}$ is $\mathscr{F}_v$ as $v\rightarrow+\infty$, whose asymptotic behavior of the least divergence term is
\begin{equation}\label{Fv}
\mathscr{F}_v\sim e^{\kappa_-v}\left(\sum_n e^{-i(-in\kappa_+)v}+\sum_n e^{-i(-in\kappa_-)v}\right)\sim e^{(\kappa_--\kappa_+)v}  \,.
\end{equation}
This is the result claimed in Section \ref{S2}.

For the stress-energy tensor (\ref{T}), the energy density $\rho$ is given by
\begin{eqnarray}
2\rho=2T_{\alpha\beta}U^\alpha U^\beta
=\frac{1}{2}\left(\psi_{,t}\psi_{,t}+\psi_{,r_*}\psi_{,r_*}\right)\left(u^t\right)^2+2\psi_{,t}\psi_{,r_*}u^t u^{r_*}
+\left(\psi_{,t}\psi_{,t}+\psi_{,r_*}\psi_{,r_*}\right)\left(u^{r_*}\right)^2         \,.
\end{eqnarray}
When $v\rightarrow+\infty$, we have
\begin{eqnarray}
\psi_{,t}&\rightarrow& i\omega X(v)+i\omega Y(u)\rightarrow X_{,-v}+Y_{,u}   \,,\nonumber\\
\psi_{,r_*}&\rightarrow& -i\omega X(v)+i\omega Y(u)\rightarrow -X_{,-v}+Y_{,u}  \,.
\end{eqnarray}
Thus, we arrive at
\begin{eqnarray}
2\rho&=&\frac{1}{2}\left[ \left( X_{,-v}+Y_{,u} \right)^2+\left( -X_{,-v}+Y_{,u} \right)^2 \right] \frac{E^2}{f^2}
+2\left( X_{,-v}+Y_{,u} \right)\left( -X_{,-v}+Y_{,u} \right)\frac{E}{f}\frac{\sqrt{E^2-f}}{f}        \nonumber\\
&&+\frac{1}{2}\left[ \left( X_{,-v}+Y_{,u} \right)^2+\left( -X_{,-v}+Y_{,u} \right)^2 \right] \frac{E^2-f}{f^2}         \,.
\end{eqnarray}
After collecting the terms inside the brackets, we obtain
\begin{eqnarray}
2\rho&=&X_{,-v}^2\left( \frac{E}{f}-\frac{\sqrt{E^2-f}}{f} \right)^2+Y_{,u}^2\left( \frac{E}{f}+\frac{\sqrt{E^2-f}}{f} \right)^2
=\mathscr{F}_v^2+\mathscr{F}_u^2        \,.
\end{eqnarray}
Considering (\ref{Fu}) and (\ref{Fv}), we finally get
\begin{equation}
\rho\sim e^{2(\kappa_--\kappa_+)v}  \,.
\end{equation}
Therefore,
\begin{equation}
\rho_\psi\sim e^{2(\kappa_- -\kappa_+) v} \rightarrow+\infty  \, , \quad v\rightarrow+\infty     \, .
\end{equation}


\begin{thebibliography}{99}



\bibitem{Penrose:1964wq}
R.~Penrose,
Phys. Rev. Lett. \textbf{14}, 57-59 (1965)
doi:10.1103/PhysRevLett.14.57


\bibitem{Hawking:1970zqf}
S.~W.~Hawking and R.~Penrose,
Proc. Roy. Soc. Lond. A \textbf{314}, 529-548 (1970)
doi:10.1098/rspa.1970.0021


\bibitem{Bardeen}
J.M.~Bardeen: Non-singular general-relativistic gravitational collapse, Proceedings of the International Conference GR5, Tbilisi, U.S.S.R. (1968), p. 174








\bibitem{Dymnikova:1992ux}
I.~Dymnikova,
Gen. Rel. Grav. \textbf{24}, 235-242 (1992)
doi:10.1007/BF00760226


\bibitem{Simpson:2019mud}
A.~Simpson and M.~Visser,
Universe \textbf{6}, no.1, 8 (2019)
doi:10.3390/universe6010008
[arXiv:1911.01020 [gr-qc]].


\bibitem{Simpson:2021dyo}
A.~Simpson and M.~Visser,
JCAP \textbf{03}, no.03, 011 (2022)
doi:10.1088/1475-7516/2022/03/011
[arXiv:2111.12329 [gr-qc]].


\bibitem{Lan:2023cvz}
C.~Lan, H.~Yang, Y.~Guo and Y.~G.~Miao,
[arXiv:2303.11696 [gr-qc]].


\bibitem{Bambi:2023try}
C.~Bambi,
Springer Singapore, 2023,
doi:10.1007/978-981-99-1596-5
[arXiv:2307.13249 [gr-qc]].





\bibitem{Modesto:2008im}
L.~Modesto,
Int. J. Theor. Phys. \textbf{49}, 1649-1683 (2010)
doi:10.1007/s10773-010-0346-x
[arXiv:0811.2196 [gr-qc]].


\bibitem{Ghosh:2014pba}
S.~G.~Ghosh,
Eur. Phys. J. C \textbf{75}, no.11, 532 (2015)
doi:10.1140/epjc/s10052-015-3740-y
[arXiv:1408.5668 [gr-qc]].







\bibitem{Frolov:2014jva}
V.~P.~Frolov,
JHEP \textbf{05}, 049 (2014)
doi:10.1007/JHEP05(2014)049
[arXiv:1402.5446 [hep-th]].


\bibitem{Frolov:2017rjz}
V.~P.~Frolov and A.~Zelnikov,
Phys. Rev. D \textbf{95}, no.12, 124028 (2017)
doi:10.1103/PhysRevD.95.124028
[arXiv:1704.03043 [hep-th]].




\bibitem{Spallucci:2008ez}
E.~Spallucci, A.~Smailagic and P.~Nicolini,
Phys. Lett. B \textbf{670}, 449-454 (2009)
doi:10.1016/j.physletb.2008.11.030
[arXiv:0801.3519 [hep-th]].


\bibitem{Koch:2014cqa}
B.~Koch and F.~Saueressig,
Int. J. Mod. Phys. A \textbf{29}, no.8, 1430011 (2014)
doi:10.1142/S0217751X14300117
[arXiv:1401.4452 [hep-th]].


\bibitem{Bodendorfer:2019jay}
N.~Bodendorfer, F.~M.~Mele and J.~M\"unch,
Class. Quant. Grav. \textbf{38}, no.9, 095002 (2021)
doi:10.1088/1361-6382/abe05d
[arXiv:1912.00774 [gr-qc]].


\bibitem{Bodendorfer:2019nvy}
N.~Bodendorfer, F.~M.~Mele and J.~M\"unch,
Phys. Lett. B \textbf{819}, 136390 (2021)
doi:10.1016/j.physletb.2021.136390
[arXiv:1911.12646 [gr-qc]].


\bibitem{Brahma:2020eos}
S.~Brahma, C.~Y.~Chen and D.~h.~Yeom,
Phys. Rev. Lett. \textbf{126}, no.18, 181301 (2021)
doi:10.1103/PhysRevLett.126.181301
[arXiv:2012.08785 [gr-qc]].




\bibitem{Hayward:2005gi}
S.~A.~Hayward,
Phys. Rev. Lett. \textbf{96}, 031103 (2006)
doi:10.1103/PhysRevLett.96.031103
[arXiv:gr-qc/0506126 [gr-qc]].



\bibitem{Nicolini:2005vd}
P.~Nicolini, A.~Smailagic and E.~Spallucci,
Phys. Lett. B \textbf{632}, 547-551 (2006)
doi:10.1016/j.physletb.2005.11.004
[arXiv:gr-qc/0510112 [gr-qc]].


\bibitem{Bonanno:2000ep}
A.~Bonanno and M.~Reuter,
Phys. Rev. D \textbf{62}, 043008 (2000)
doi:10.1103/PhysRevD.62.043008
[arXiv:hep-th/0002196 [hep-th]].


\bibitem{Gambini:2008dy}
R.~Gambini and J.~Pullin,
Phys. Rev. Lett. \textbf{101}, 161301 (2008)
doi:10.1103/PhysRevLett.101.161301
[arXiv:0805.1187 [gr-qc]].




\bibitem{Modesto:2004xx}
L.~Modesto,
Phys. Rev. D \textbf{70}, 124009 (2004)
doi:10.1103/PhysRevD.70.124009
[arXiv:gr-qc/0407097 [gr-qc]].






\bibitem{Nicolini:2008aj}
P.~Nicolini,
Int. J. Mod. Phys. A \textbf{24}, 1229-1308 (2009)
doi:10.1142/S0217751X09043353
[arXiv:0807.1939 [hep-th]].












\bibitem{Lewandowski:2022zce}
J.~Lewandowski, Y.~Ma, J.~Yang and C.~Zhang,
Phys. Rev. Lett. \textbf{130}, no.10, 101501 (2023)
doi:10.1103/PhysRevLett.130.101501
[arXiv:2210.02253 [gr-qc]].


\bibitem{Yang:2022btw}
J.~Yang, C.~Zhang and Y.~Ma,
Eur. Phys. J. C \textbf{83}, no.7, 619 (2023)
doi:10.1140/epjc/s10052-023-11800-8
[arXiv:2211.04263 [gr-qc]].


\bibitem{Zhang:2023okw}
C.~Zhang, Y.~Ma and J.~Yang,
[arXiv:2302.02800 [gr-qc]].


\bibitem{EventHorizonTelescope:2019dse}
K.~Akiyama \textit{et al.} [Event Horizon Telescope],
Astrophys. J. Lett. \textbf{875}, L1 (2019)
doi:10.3847/2041-8213/ab0ec7
[arXiv:1906.11238 [astro-ph.GA]].

\bibitem{Simpson:1973ua}
M.~Simpson and R.~Penrose,
Int. J. Theor. Phys. \textbf{7}, 183-197 (1973)
doi:10.1007/BF00792069

\bibitem{Mcnamara:1977}
Mcnamara, John. (1977). 
Royal Society of London Proceedings Series A. 358. 499-517. 10.1098/rspa.1978.0024.


\bibitem{Mcnamara:1978}
McNamara John Michael 1978
Proc. R. Soc. Lond. A364121-134


\bibitem[Chandrasekhar and Hartle(1982)]{1982RSPSA.384..301C}
S.~Chandrasekhar,  and J.~B.~Hartle, : 1982, {\it Proceedings of the Royal Society of London Series A} {\bf 384}, 301. doi:10.1098/rspa.1982.0160.




\bibitem{Brown:2011tv}
E.~G.~Brown, R.~B.~Mann and L.~Modesto,
Phys. Rev. D \textbf{84}, 104041 (2011)
doi:10.1103/PhysRevD.84.104041
[arXiv:1104.3126 [gr-qc]].


\bibitem{Iofa:2022dnc}
M.~Z.~Iofa,
J. Exp. Theor. Phys. \textbf{135}, no.5, 647-654 (2022)
doi:10.1134/S1063776122110048
[arXiv:2206.09460 [gr-qc]].


\bibitem{Carballo-Rubio:2021bpr}
R.~Carballo-Rubio, F.~Di Filippo, S.~Liberati, C.~Pacilio and M.~Visser,
JHEP \textbf{05}, 132 (2021)
doi:10.1007/JHEP05(2021)132
[arXiv:2101.05006 [gr-qc]].



\bibitem{Kazakov:1993ha}
D.~I.~Kazakov and S.~N.~Solodukhin,
Nucl. Phys. B \textbf{429}, 153-176 (1994)
doi:10.1016/S0550-3213(94)80045-6
[arXiv:hep-th/9310150 [hep-th]].









\bibitem{R. Penrose:1974}
R.~Penrose, 1974,
Cambridge University Press, Symposium - International Astronomical Union, vol. 64, pp. 82-91.



\bibitem{Brown:2010csa}
E.~Brown, R.~B.~Mann and L.~Modesto,
Phys. Lett. B \textbf{695}, 376-383 (2011)
doi:10.1016/j.physletb.2010.11.035
[arXiv:1006.4164 [gr-qc]].




\bibitem[Chandrasekhar(1983)]{1983mtbh.book.....C} S.~Chandrasekhar, \ 1983, The International Series of Monographs on Physics, Oxford: Clarendon Press, 1983


\bibitem{Ashtekar:2006rx}
A.~Ashtekar, T.~Pawlowski and P.~Singh,
Phys. Rev. Lett. \textbf{96}, 141301 (2006)
doi:10.1103/PhysRevLett.96.141301
[arXiv:gr-qc/0602086 [gr-qc]].

\bibitem{Kelly:2020uwj}
J.~G.~Kelly, R.~Santacruz and E.~Wilson-Ewing,
Phys. Rev. D \textbf{102}, no.10, 106024 (2020)
doi:10.1103/PhysRevD.102.106024
[arXiv:2006.09302 [gr-qc]].


\bibitem{Ori:1991zz}
A.~Ori,
Phys. Rev. Lett. \textbf{67}, 789-792 (1991)
doi:10.1103/PhysRevLett.67.789


\bibitem{Poisson:1990eh}
E.~Poisson and W.~Israel,
Phys. Rev. D \textbf{41}, 1796-1809 (1990)
doi:10.1103/PhysRevD.41.1796


\bibitem{Price:1972pw}
R.~H.~Price,
Phys. Rev. D \textbf{5}, 2439-2454 (1972)
doi:10.1103/PhysRevD.5.2439

\bibitem{Han:2023wxg}
M.~Han, C.~Rovelli and F.~Soltani,
Phys. Rev. D \textbf{107}, no.6, 064011 (2023)
doi:10.1103/PhysRevD.107.064011
[arXiv:2302.03872 [gr-qc]].


\bibitem{Penrose:1969pc}
R.~Penrose,
Riv. Nuovo Cim. \textbf{1}, 252-276 (1969)
doi:10.1023/A:1016578408204

\bibitem{Carballo-Rubio:2019fnb}
R.~Carballo-Rubio, F.~Di Filippo, S.~Liberati and M.~Visser,
Phys. Rev. D \textbf{101}, 084047 (2020)
doi:10.1103/PhysRevD.101.084047
[arXiv:1911.11200 [gr-qc]].

\bibitem{Carballo-Rubio:2022kad}
R.~Carballo-Rubio, F.~Di Filippo, S.~Liberati, C.~Pacilio and M.~Visser,
JHEP \textbf{09}, 118 (2022)
doi:10.1007/JHEP09(2022)118
[arXiv:2205.13556 [gr-qc]].



\end{thebibliography}

\end{document}